\documentstyle[11pt,moriond,epsfig]{article}

\def\Journal#1#2#3#4{{#1} {\bf #2}, #3 (#4)}

\def\PLB{{\em Phys. Lett.}  B}

\def\EPJ{{\em Eur. Phys. J.} C}

\def\be{\begin{equation}}
\def\ee{\end{equation}}
\def\bea{\begin{eqnarray}}
\def\eea{\end{eqnarray}}

\newcommand{\epem}{\mbox{$\mbox{\rm e}^{+}\mbox{\rm e}^{-}$}}

\newcommand{\lplm}{\mbox{$l^{+}l^{-}$}}
\newcommand{\Zz}{\mbox{Z$^0$}}

\newcommand{\WW}{\mbox{$\mbox{\rm W}^{+}\mbox{\rm W}^{-}$}}

\newcommand{\qq}{\mbox{\rm q}\mbox{$\overline{\mbox{\rm q}}$}}

\newcommand{\lnu}{\mbox{$l\overline{\nu}_{l}$}}
\newcommand{\lnup}{\mbox{$l^{+}\nu$}}
\newcommand{\lnum}{\mbox{$l^{-}\overline{\nu}$}}
\newcommand{\enu}{\mbox{e$\overline{\nu}_{e}$}}
\newcommand{\mnu}{\mbox{$\mu\overline{\nu}_{\mu}$}}
\newcommand{\tnu}{\mbox{$\tau\overline{\nu}_{\tau}$}}
\newcommand{\qqqq}{\qq\qq}
\newcommand{\qqln}{\qq\,\lnu}
\newcommand{\qqen}{\qq\enu}
\newcommand{\qqmn}{\qq\mnu}
\newcommand{\qqtn}{\qq\tnu}

\newcommand{\WWqqln}{\WW$\rightarrow$\qqln}
\newcommand{\WWqqqq}{\WW$\rightarrow$\qqqq}
\newcommand{\WWqqen}{\WW$\rightarrow$\qqen}
\newcommand{\WWqqmn}{\WW$\rightarrow$\qqmn}
\newcommand{\WWqqtn}{\WW$\rightarrow$\qqtn}
\newcommand{\lnln}{\lnup\lnum}
\newcommand{\WWlnln}{\WW$\rightarrow$\lnln}
\newcommand{\Zqq}{\Zz/$\gamma\rightarrow$\qq}

\newcommand{\Mw}{$M_{\mathrm{W}}$}

\newcommand{\Wenu}{\mbox{\epem$\rightarrow \mbox{\rm W}$\enu}}
\newcommand{\ZZqqll}{(\Zz/$\gamma$)$^{*}$(\Zz/$\gamma$)$^{*}\rightarrow$\qq\lplm}

\begin{document}
\vspace*{4cm}
\title{W mass measurement at LEP}

\author{ T. Saeki }

\address{International Center for Elementary Particle Physics, \\
University of Tokyo \\
7-3-1 Hongo, Bunkyo-ku, Tokyo 113-0033, Japan}

\maketitle\abstracts{
In 1998, the four experiments of LEP, i.e. ALEPH, DELPHI, L3 and OPAL,
collected data of about 175 pb$^{-1}$ per experiment at the center-of-mass 
energy of 189 GeV. Using these data, the mass of W boson was directly 
measured by reconstructing the decay products of two W bosons from
the e+e- collisions. The W mass measurement was combined personally
with the results obtained from data at 161, 172, and 183 GeV.
This yielded the private LEP2 average of \Mw $=80.350 \pm 0.056$ GeV.
}

\section{Introduction}
%\subsection{Producing the Hard Copy}\label{subsec:prod}

In June 1996, the LEP2 started with the center-of-mass energy (CME) of 161 GeV,
just above the threshold of pair-production of W bosons.
This allowed the four experiments of LEP, i.e. ALEPH, DELPHI, L3 and OPAL, 
to collect data of about 10 pb$^{-1}$ per experiment and to measure the mass 
of W boson from cross-section measurement of WW events.
In October 1996, and in 1997 and 1998, the CME's were raised to 
significantly above the threshold, 172, 183 and 189 GeV, and
the recorded data per experiment were about 10, 55 and 175 pb$^{-1}$,
respectively. Using these data, the mass of W boson was directly 
measured by reconstructing decay products of W boson pair.

\section{Selection}

WW events are produced through three doubly resonant diagrams 
($s$-channel $\gamma$ and Z$^0$ exchange and $t$-channel 
$\nu$ exchange), called "CC03 diagrams", where
each W can decay into quark pair or lepton-nutorino pair.
This leads to the classification of WW events into three channels,
i.e. fully hadronic, semileptonic, and fully leptonic channels.
WW events are selected 
with good efficiency and high purity in the analysis,
utilising corresponding event-topology to the three channels.

Hadronic \WWqqqq\ decays comprise 46\% of the total \WW\ cross-section.
The typical final state of the fully hadronic events is specified 
by four hadronic jets whose energy sum is consistent with the 
center-of-mass energy.
The background is dominated by electron-positron annihilation
to $q\overline{q}(\gamma)$.
Semileptonic final states, \WWqqln, are expected to
comprise 44\% of \WW\ decays.
\WWqqen\ and \WWqqmn\ events are characterised by two 
well-separated hadronic jets, a high momentum lepton and 
sizable missing momentum due to the unobserved neutrino. 
The signature for 
\WWqqtn\ events is two well separated jets from
the hadronic W decay and one low multiplicity jet typically consisting of
one or three tracks due to the decay of tau. 
The dominant backgrounds are \Zqq\ and four-fermion processes such as 
\Wenu\ and \ZZqqll.
Fully leptonic \WWlnln\ decays comprise 10\% of the total \WW\ cross-section.
The typical fully leptonic WW events consist of
two acoplanar energetic leptons with significant missing energy 
in detectors. Typical efficiency and purity of WW event selection and
expected and observed numbers of events in each channel,
from ALEPH at 189 GeV, are shown in table \ref{tab:eff}.

\begin{table}[htbp]
\caption{
   WW event selection of ALEPH at 189 GeV.
   \label{tab:eff}}
   \vspace{0.4cm}
 \begin{center}
  \begin{tabular}{|l||c|c|c|c|} \hline
           & Efficiency (\%) 
             & ~~Purity (\%)~~ 
               & Exp. N evts 
                 & Obs. N evts \\ \hline
    \qqqq  &  71.4                
             &  84.7             
               &  1173             
                 &  1068       \\ 
    \qqen  &  81.3               
             &  92.7             
               &  371           
                 &  358        \\ 
    \qqmn  &  84.1               
             &  96.6             
               &  365           
                 &  363        \\
    \qqtn  &  41.5               
             &  95.4             
               &  176           
                 &  159        \\ \hline
  \end{tabular}
 \end{center}
\end{table} 

\section{Extraction of W mass}

As to the actual procedure to measure \Mw\ from W-pair production,
two methods are advocated. One procedure requires a measurement of 
the total W-pair cross-section close to the threshold,
where the size of $\sigma_{\mathrm tot}$ is most sensitive to
the W mass.  This method is adopted at the CME of 161 GeV
and the result of W mass measurements is $80.40 \pm 0.22$ GeV 
~\cite{MWSIG1,MWSIG2} with combining four experiments. 

The other one is called the direct mass reconstruction method,
which is adopted at the CME's of 172, 183 and 189 GeV.
In this method, the measurement of the W mass can be made by 
direct reconstruction of the invariant mass of the fermion pairs 
from each W decay, using a kinematic fit technique with some constraints.
Incorporating the constraints of energy and momentum conservation
into a kinematic fit significantly improves the invariant mass
resolution and is adopted by all experiments.
Specific combinations of additional constraints and techniques, for example, 
a constraint of equal mass of two W bosons, a technique of 
beam energy rescaling and so on, are employed by some experiments.

\begin{figure}[htp]
\begin{center}
\epsfig{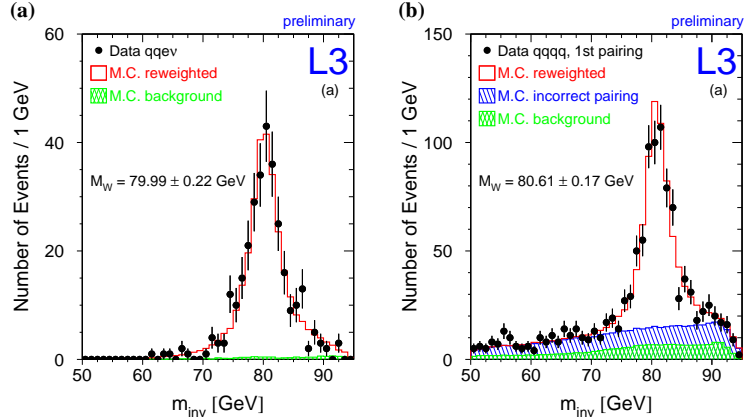}
\caption[]{\small The invariant mass distributions from L3 at 189 GeV.
           \label{fig:minv}}
\end{center}
\end{figure}

Events of the fully hadronic and semileptonic decay channels 
are used in the analysis. 
In the fully hadronic channel, four jets in an event can be 
divided into two di-jets in three different ways.
It is not obvious which of these partitions is correct and
so this ambiguity leads to a combinatorial background.
Four experiments employed different pairing schemes 
to optimise the sensitivity to the W mass.
In a \WWqqln\ event ($l=$e or $\mu$), a kinematic fit is performed 
including two jets and one lepton, imposing the constraints 
mentioned above.
For \WWqqtn\ events, ALEPH employed a kinematic fit, 
L3 and OPAL utilised a technique of beam energy rescaling.
Fig. \ref{fig:minv} shows the distributions
of the reconstructed invariant mass from L3 for (a) semileptonic 
and (b) fully hadronic channels, where
a larger amount of background in fully hadronic channel than
the semileptonic channel is due to the above mentioned jet-pairing
combinatorial background.

The invariant mass distributions obtained from the event sample 
have a Breit-Wigner like shape but distorted due to several effects such 
as phase space restrictions, detector resolution, initial state radiation,
background contamination, selection algorithms, etc... 
A possible way of extracting \Mw\ is
fitting directly to the data the invariant mass distribution 
predicted by the Monte Carlo (MC) including all distortions.
In this method, MC events for some specific input values of \Mw$^{MC}$
are generated, and by reweighting technique,
a MC sample with an arbitrary input value of \Mw$^{MC}$ is produced. 
Then one can find the best matching MC sample to the data, 
where that \Mw$^{MC}$ yields the measurement of \Mw.
ALEPH, L3 and OPAL employed this method to extract 
the W mass~\cite{ALO172,ALO183}. 

On the other hand, DELPHI developed a different method
in which the information on the W mass is extracted from
the likelihood of observing each individual event~\cite{D172}.
%\begin{eqnarray}
%   {\mathcal{L}} (M_{\mathrm W}) =  P \left(
%    \int_{0}^{E_b} G(m|m_{\mathrm f}) BW(m|M_{\mathrm W}) PS(m) dm
%    \right) + (1-P) p_b (m_{\mathrm f}),
%\end{eqnarray}
The event-by-event likelihood as a function of \Mw\ is calculated by
the convolution of a Gaussian resolution function with a mean of 
fitted invariant mass and a width of its error, and a relativistic 
Breit-Wigner function of \Mw\ including the phase space effect, 
taking into account the efficiency, purity, background, 
all jet-pairing combinations in \qqqq\ channel and so on.
The combined likelihood for observing all the events is expressed as 
the product of all the event-by-event likelihoods.
The maximum of this combined likelihood then yields the
measurement of the W mass.

\section{Results}

\begin{table}[htbp]
  \caption{
   Official and private averages of LEP \Mw\ measurements 
   by direct reconstruction method.
   \label{tab:result}}
   \vspace{0.4cm}
  \begin{center}
    \begin{tabular}{|l||c|c|} \hline
      W mass results   & Mw $\pm$ (stat.) $\pm$ (syst.) & Data  \\
      ~~~~~(GeV)       & $\pm$ (FSI) $\pm$ (LEP)        &       \\ \hline
      Official average & $80.368 \pm 0.044 \pm 0.039$   & A+L 172-189 GeV \\
                       & $\pm 0.023 \pm 0.018$          
                       & D+O 172-183 GeV \\ \hline
      Private average  & $80.347 \pm 0.036 \pm 0.036$   & A+D+L+O     \\
                       & $\pm 0.020 \pm 0.017$          & 172-189 GeV \\ \hline
    \end{tabular}
  \end{center}
\end{table} 

The official average of measured \Mw's from ALEPH and L3 using 
172 - 189 GeV data and DELPHI and OPAL using 172 - 183 GeV data
is shown in table \ref{tab:result}.
The first, second, third and fourth errors are the statistical and systematic
uncertainties, and the uncertainties from the final state interactions (FSI) 
and the LEP beam energy (LEP), respectively. 
During this conference, on 23 March 1999, the results of DELPHI and OPAL
at 189 GeV were approved and I privately averaged all these results. 
This private combined results of LEP four experiments is also shown 
in table \ref{tab:result}. The systematic error and uncertainty
from FSI will be mentioned in next section.

Recently ALEPH released a new analysis in which \Mw\ is extracted 
from W $\rightarrow$ \lnu\ decays using 57 pb$^{-1}$ data at 183 GeV. 
The result is \Mw $= 80.142 \pm 0.192 \pm 0.089$ GeV combining 
results from \lnln\ and \qqln\ channels, where the first and
second errors are statistical and systematic uncertainties, respectively.

\section{Systematic errors}

Typical systematic errors on \Mw\ measurement from OPAL at 183 GeV are 
shown in table \ref{tab:sys} for \qqqq, \qqln\ and combined results. 
The uncertainties of \Mw\ measurement are from LEP beam energy precision, 
theoretical uncertainty of initial state radiation, 
hadronization model dependence in MC's,
effects of not including interference terms between CC03 and other
four-fermion diagrams in reweighting procedure, 
detector effects, reweighting fit procedure,
uncertainties of normalization and shape of background distributions in 
MC samples, finite statistics of used MC samples, 
Bose-Einstein (BE) correlations and Colour Reconnection (CR) effects.

BE correlations and CR effects are simply called 
final state interactions and then abbreviated as FSI.
Uncertainty from BE correlations happens only in \qqqq\ channel,
because BE correlations between decay products from different W's
might distort the invariant mass spectrum.
Two MC samples with and without this effect are compared,
and the difference of measured \Mw's is assigned as systematic error.
Uncertainty from CR effects also happens only in \qqqq\ channel.
In normal MC's, fragmentation is implemented only within each W, 
but fragmentation between two W's might distorts the invariant mass spectrum.
Various MC models including CR effects are checked using 183 GeV data
and the Ellis-Geiger model VNI was excluded~\cite{O183CR}. 
In OPAL, ARIADNE model is used to assign the systematic 
error by comparing it with a normal MC sample~\cite{ALO183}.

\begin{table}[htbp]
  \caption{
    Summary of the systematic uncertainties from OPAL at 183 GeV.
    \label{tab:sys}}
   \vspace{0.4cm}
  \begin{center}
    \begin{tabular}{|l||c|c|c|} \hline
      Systematic errors &\multicolumn{3}{|c|}{ \Mw }      \\
        ~~~~~~~~(MeV) 
      & \qqqq & \qqln & comb.                       \\ \hline
      Beam Energy                & 22  & 22 & 22    \\
      Initial State Radiation    & 10  & 10 & 10    \\
      Hadronization              & 21  & 21 & 16    \\
      Four-fermion               & 30  & 28 & 21    \\
      Detector Effects           & 38  & 31 & 26    \\
      Fit Procedure              & 15  & 15 & 15    \\
      Background                 & 25  & 10 & 10    \\
      MC statistics              & 15  & 15 & 11    \\ \hline
      Sub-total                  & 67  & 58 & 49    \\
      Bose-Einstein Correlations & 32  &  0 & 11    \\ 
      Colour Reconnection        & 49  &  0 & 16    \\ \hline\hline
      Total systematic error     & 89  & 58 & 53    \\ \hline
    \end{tabular}
  \end{center}
\end{table} 

\section{Conclusions}

The four experiments of LEP collected data successfully at the CME's of
161, 172, 183 and 189 GeV. The official average of \Mw\ using the data 
of (A+L 172 - 189 GeV) and (D+O 172 - 183 GeV) is 
$80.368 \pm 0.044$(stat.) $\pm 0.039$(syst.) $\pm 0.023$(FSI) 
$\pm 0.018$(LEP) GeV. 
If combining it with results at 161 GeV, the official LEP2 average is \Mw\ 
$= 80.370 \pm 0.063$ GeV.
Including DELPHI and OPAL results at 189 GeV, the private average
of \Mw, where the data is (A+D+L+O 172 -189 GeV), is 
$80.347 \pm 0.036$(stat.) $\pm 0.036$(syst.) $\pm 0.020$
(FSI) $\pm 0.017$(LEP) GeV. If this is combined with results at 161 GeV,
the private LEP2 average is 

\begin{center} 
\Mw\ $= 80.350 \pm 0.056$ GeV.
\end{center} 

Because the systematic error on \Mw\ measurement is now 
less than the statistical error,
more studies are needed on sytematic errors from 
BE correlations, CR effects, LEP beam energy, detector effects and so on.
Also important is utilizing W $\rightarrow$ \lnu\ decays to reduce
the error on \Mw\ measurement.

\begin{figure}[htp]
\begin{center}
\epsfig{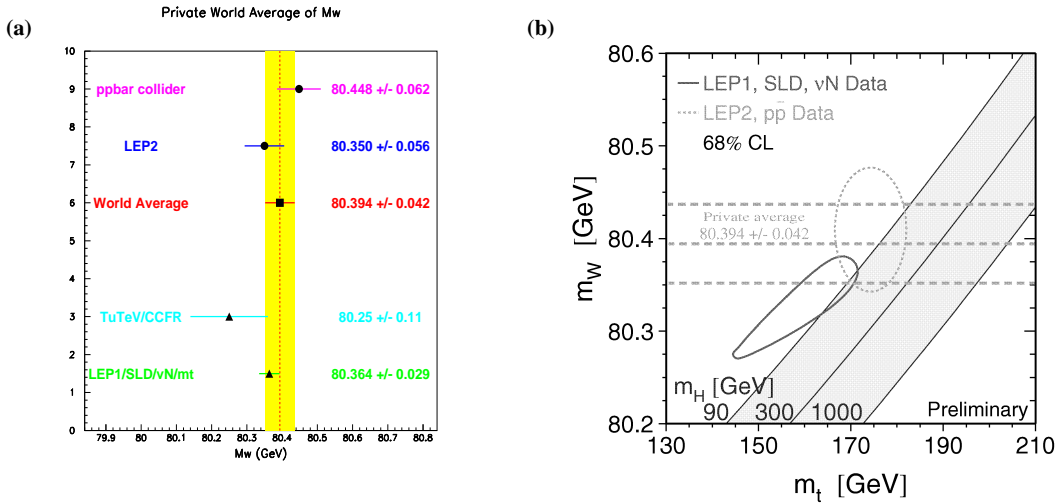}
\end{center}
\caption[]{
   (a) the private world average of W mass direct measurements 
   and comparison with indirect W mass measurements,
   and (b) the comparison of the indirect measurements of $m_{\mathrm W}$
   and $m_{\mathrm t}$ (LEP+SLD+$\nu$N data) (solid contour) and
   the direct measurements (Tevatron and LEP2 official averages) 
   (dashed contour). In both cases the 68\% C.L. contours are plotted.
   \label{fig:wa_mwmt}}
\end{figure}

The private world average of W mass direct measurements 
combining the private LEP2 average and results of direct measurements
from Tevatron~\cite{PPBAR} is shown in figure \ref{fig:wa_mwmt} (a).
In the same figure, the private world average of 
\Mw\ $=80.394 \pm 0.042$ GeV is compared with 
the indirect W mass measurements from TuTeV/CCFR and 
LEP1/SLD/$\nu$N/$m_{\mathrm t}$~\cite{SMFIT}.
The direct and indirect measurements are in agreement within errors.
The comparison of the indirect measurements of $m_{\mathrm W}$
and $m_{\mathrm t}$ (LEP+SLD+$\nu$N data)~\cite{SMFIT} (solid contour) and
the direct measurements (Tevatron~\cite{PPBAR} and LEP2 official averages)
(dashed contour) is shown in figure \ref{fig:wa_mwmt} (b). 
Also shown in the figure are the Standard Model relationship 
for the masses as a function of the Higgs mass, 
and the private world average of W mass in three dashed lines 
with a width of the measurement error.

\vspace{1cm}

%\begin{figure}[htp]
%\begin{center}
%\epsfig{figure=figure/mwplot_wa1.eps,width=8cm}
%\caption[]{\small \it The 95\% CL upper bound
%\label{fig_1}}
%\end{center}
%\end{figure}

%\begin{figure}[htp]
%\begin{center}
%\epsfig{figure=figure/morion_p24+.eps,width=8cm}
%%\epsfig{figure=figure/m99_mt_mw_contours.eps,width=8cm}
%\caption[]{\small \it The 95\% CL upper bound
%\label{fig_2}}
%\end{center}
%\end{figure}

\section*{Acknowledgments}

I would like to thank the CERN SL Division for their successful 
operation of accelerator and the four LEP collaborations 
for providing me with information.
Especially the suport of the LEP EW working group is thankfully 
acknowledged.

\end{document}